\documentclass[twocolumn]{aastex63}
\pdfoutput=1 %for arXiv submission
\usepackage{appendix}
\usepackage{amsmath,amstext}
\usepackage[T1]{fontenc}
\usepackage[figure,figure*]{hypcap}

\newcommand{\mos}{\,m\,s$^{-1}$}
\newcommand{\kms}{\,km\,s$^{-1}$}

\graphicspath{{./}{figures/}}

\received{?}
\revised{?}
\accepted{?}
\submitjournal{AAS}

\shorttitle{Spin-Orbit Alignment in K2-232}
\shortauthors{Wang et al.}
\begin{document}

\title{The Aligned Orbit of the Eccentric Warm Jupiter K2-232\,b}

\author[0000-0002-7846-6981]{Songhu Wang}
\affil{Department of Astronomy, Indiana University, Bloomington, IN 47405}
\email{sw121@iu.edu}

\author[0000-0002-4265-047X]{Joshua N.\ Winn}
\affiliation{Department of Astrophysical Sciences, Princeton University, 4 Ivy Lane, Princeton, NJ 08544}

\author[0000-0003-3216-0626]{Brett C.\ Addison}
\affil{University of Southern Queensland, Centre for Astrophysics, West Street, Toowoomba, QLD 4350 Australia} 

\author[0000-0002-8958-0683]{Fei Dai}
\affiliation{Division of Geological and Planetary Sciences 1200 E California Blvd, Pasadena, CA 91125}

\author[/0000-0002-7670-670X]{Malena Rice}
\affil{Department of Astronomy, Yale University, New Haven, CT 06511}

\author[0000-0002-6153-3076]{Bradford Holden}
\affil{UCO/Lick Observatory, Department of Astronomy and Astrophysics, University of California at Santa Cruz,Santa Cruz, CA 95064}

\author[0000-0002-0040-6815]{Jennifer A. Burt}
\affil{Jet Propulsion Laboratory, California Institute of Technology, 4800 Oak Grove drive, Pasadena CA 91109}

\author[0000-0002-0376-6365]{Xian-Yu Wang}
\affiliation{National Astronomical Observatories, Chinese Academy of Sciences, Beijing 100012, China}
\affiliation{University of Chinese Academy of Sciences, Beijing, 100049, China}

\author[0000-0003-1305-3761]{R. Paul Butler}
\affil{ Earth and Planets Laboratory, Carnegie Institution for Science, 5241 Broad Branch Road, NW, Washington, DC 20015}

\author[0000-0001-7177-7456]{Steven S.\ Vogt}
\affil{UCO/Lick Observatory, Department of Astronomy and Astrophysics, University of California at Santa Cruz,Santa Cruz, CA 95064}

\author[0000-0002-3253-2621]{Gregory Laughlin}
\affil{Department of Astronomy, Yale University, New Haven, CT 06511}



\begin{abstract}
\noindent 

Measuring the obliquity distribution of stars hosting warm Jupiters may help us to understand
the formation of close-orbiting gas giants.  Few
such measurements have been performed due to practical difficulties in scheduling
observations of the relatively infrequent and long-duration transits of warm Jupiters.
Here, we report a measurement of the Rossiter-McLaughlin effect for
K2-232\,b, a warm Jupiter %($M_{\rm P}=0.39$ $M_{\rm Jup}$) 
on an $11.17$-day orbit with an eccentricity of $0.26$.
The data were obtained with the Automated Planet Finder during two
separate transits.
The planet's orbit appears to be well-aligned with the spin axis of the host star,
with a projected spin-orbit angle of
$\lambda = -11.1\pm 6.6^{\circ}$.
Combined with the other available data, we find that high obliquities are almost
exclusively associated with planets that either have an orbital separation greater than 10
stellar radii or orbit stars with effective temperatures hotter than 6{,}000\,K.
This pattern suggests that the obliquities of the closest-orbiting giant
planets around cooler stars have been damped by
tidal effects.

%This measurement provides further evidence that while the closest-in planets orbiting cool stars tend to have both circular and well-aligned orbits, consistent with expectations with tidal damping, planets on even slightly longer orbits are not constrained in the same way.

\end{abstract}



\section{Introduction} \label{sec:intro}

In the Solar System, the orbits of the major planets are all aligned with the net angular momentum vector of the Solar System planets to within a few degrees, and the Sun's equator is tilted by only $6^\circ$ \citep{Souami2012} relative to the invariable plane. The coplanarity of the Solar System was part of the original evidence leading to the proposal that
the planets formed within a flat disk surrounding the Sun \citep{Kant1755, Laplace1796}.

\begin{figure}
\includegraphics[width = 1.0\columnwidth]{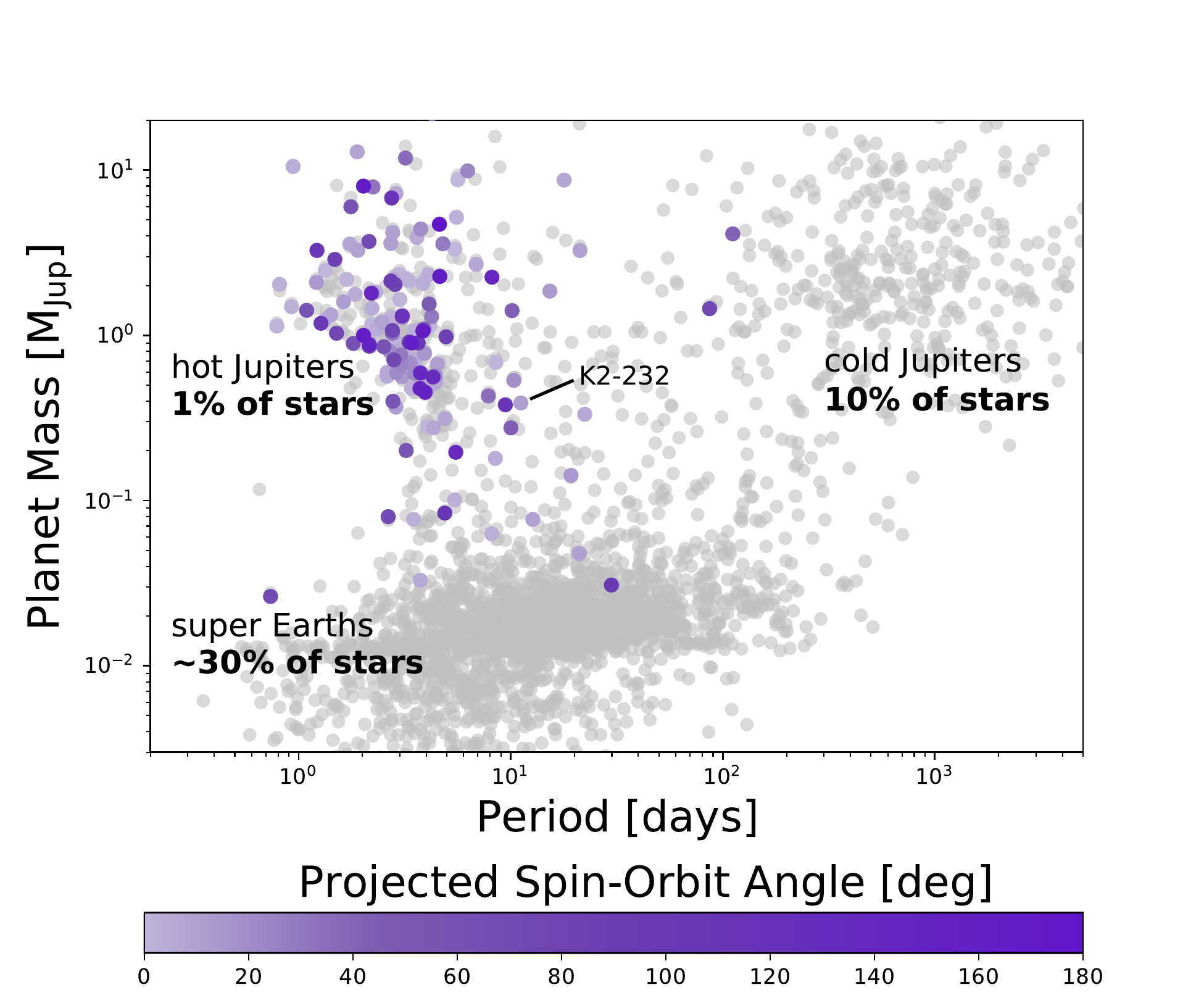}
\caption{An up-to-date mass-period diagram of currently known exoplanets.
Planets with Rossiter-McLaughlin (including Doppler tomography) measurements drawn from the TEPCat catalog \citep{Southworth2011} are shown as points color-coded by their observed spin-orbit angles, while planets without stellar obliquity measurements are depicted as gray dots. The majority of planets with existing Rossiter-McLaughlin measurements are hot Jupiters, which span a wide range of stellar obliquities. In this work, we have expanded the list of spin-orbit measurements to include a new warm Jupiter system: K2-232.}
\label{rm_dis}
\end{figure}

In contrast to the picture presented by our Solar System, observations of the Rossiter–McLaughlin effect \citep{rossiter1924detection, mclaughlin1924some} occurring during exoplanetary transits have revealed that a considerable fraction of hot Jupiters are on orbits that are misaligned with the equators of their host stars \citep{Winn2010, Albrecht2012}. The origins of large spin-orbit misalignments for hot Jupiters are still unclear. The current theoretical explanations fall into two categories: 
\begin{enumerate}

\item High-eccentricity migration, in which dynamical interactions tilt the orbit
of the planet away from its initial plane.  These theories invoke phenomena such as
planet-planet scattering \citep{Rasio1996, Ford2008},
Lidov-Kozai cycles with tidal friction \citep{Wu2003, Fabrycky2007, Naoz2016},
and secular interactions \citep{Wu2011, Petrovich2015}. 

\item Obliquity excitation via mechanisms that are \textit{unrelated}
to planet migration. These include chaotic star formation \citep{Bate2010},
stochastic internal gravity waves \citep{Rogers2012},
magnetic torques between a young star and its protoplanetary disk \citep{Lai2011},
and gravitational torques from distant companions \citep{Batygin2011, Storch2014}.
\end{enumerate}

Observations of ``warm Jupiters'' --- giant planets with orbital periods longer than about 10 days --- may be helpful in evaluating these formation scenarios. Unlike hot Jupiters, longer-period warm Jupiters experience relatively weak tidal interactions and can only form through high-eccentricity migration under certain conditions \citep{dong2013warm}. Therefore, if high obliquities are indeed associated with high-eccentricity migration, then large spin-orbit misalignments should be confined to hot Jupiters, while warm Jupiters should have orbits that are roughly aligned with the stellar equators. If the true explanation is found in the second category of theories,
spin-orbit misalignments should occur not only in hot Jupiter systems, but also in a broader class of planetary systems, including warm Jupiters.
%They should also be found in systems with low or moderate
%eccentricities, which would be challenging to explain via
%high-eccentricity migration \citep{Petrovich2016}.
%Orbital eccentricities are easier to measure for warm Jupiters
%than the more prevalent smaller planets.

Furthermore, the distribution of spin-orbit angles of warm Jupiter hosts might
be easier to interpret than that of hot Jupiter hosts
because warm Jupiters are unlikely to have influenced the stellar rotation
through tides or other proximity effects.

An effective method for measuring or placing bounds on a star's obliquity
is to observe the Rossiter-McLaughlin effect, which
requires spectroscopy throughout a transit. This requirement helps to explain
why there are relatively few observations of the Rossiter-McLaughlin effect
for warm Jupiters.  Compared to hot Jupiters, the warm Jupiters have
lower transit probabilities.  
Even for the warm Jupiters that do transit, the transits are less frequent, and the transit durations are longer.
This makes it challenging to schedule
transit observations from a single observatory, and it renders the observations, which require several consecutive hours of reliable data, more vulnerable to disruptions by bad weather.
Therefore, while stellar obliquities of warm Jupiters may
provide important clues to understand the diverse architectures of exoplanet systems,
the existing measurements are sparse.

Here, we present observations of the Rossiter-McLaughlin effect
for K2-232\,b, a warm Jupiter of mass $0.39~M_{\rm Jup}$ and radius $1.06~R_{\rm Jup}$ on an 11.17-day orbit with an eccentricity of 0.26
\citep{Brahm2018, Yu2018}.
In what follows, we describe our observations
(\S 2), the parameterized model we used to determine the spin-orbit angle
(\S 3), and the possible implications (\S 4). 

\section{Observations}
%starname in vst: k2apf8
%oct transit: l27.logsheet11 rl27.4099-4120 
%jan transit: rl29.logsheet2, rl29.386-409

\begin{figure}
   \includegraphics[width = 1.0\columnwidth]{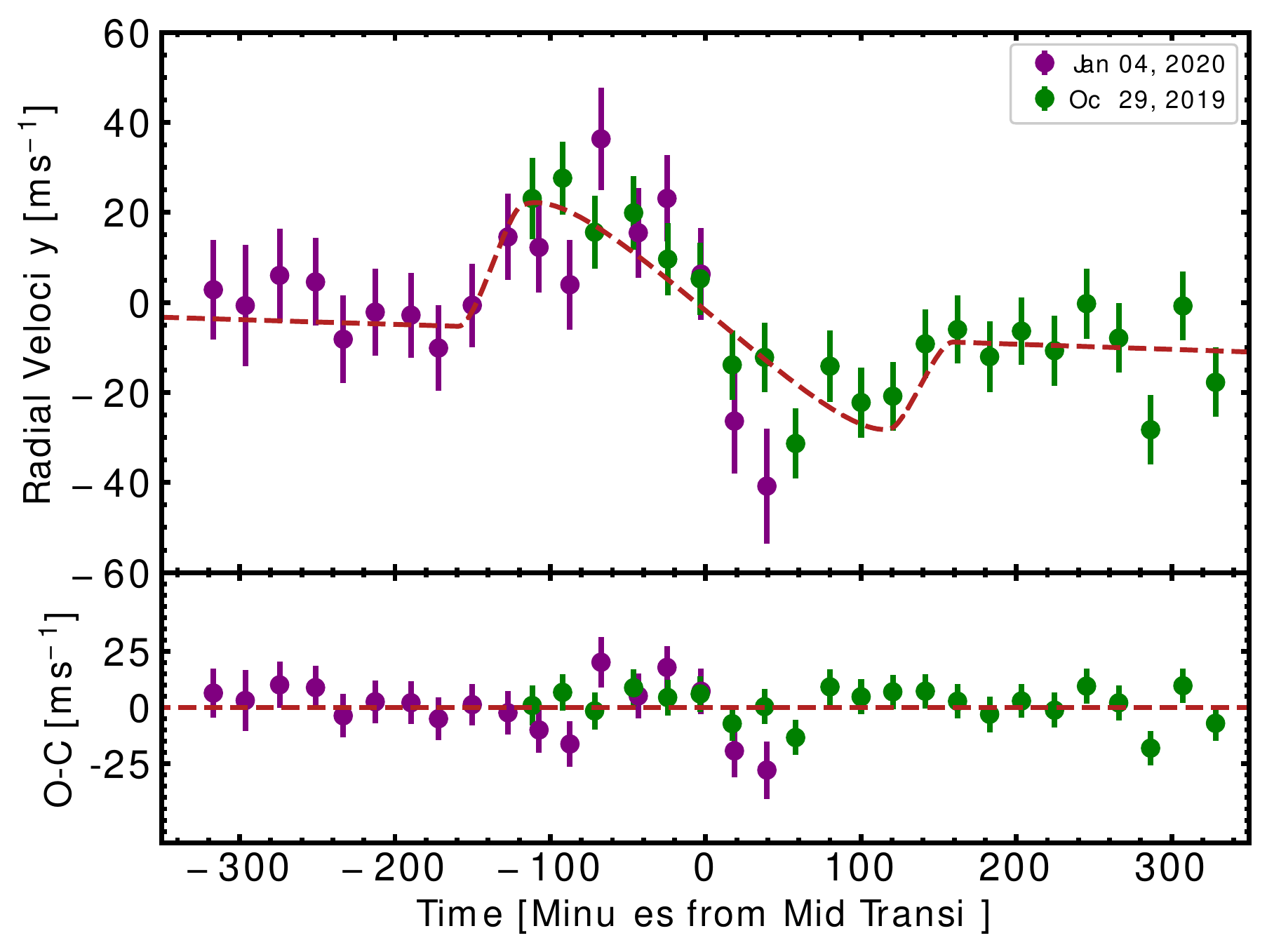}
    \caption{Spectroscopic radial velocities of K2-232 measured with the APF, as a function of orbital phase (minutes from mid-transit) along with the best-fitting Rossiter-McLaughlin model (red-dashed line). The radial velocity offsets in each of the datasets have been removed prior to fitting the combined and phased data. The green and maroon points are the radial velocities from the 29 October 2019 and 4 January 2020 transit observations, respectively. The lower panel shows the residuals between the observed data and the best-fitting model. The small structures remaining in the residuals are likely caused by stellar noise and variations in atmospheric extinction due to the presence of clouds at the end of night on 4 January 2020}.
    \label{fig:combined_transit}
\end{figure}

\begin{figure}
   \includegraphics[width = 0.9 \columnwidth]{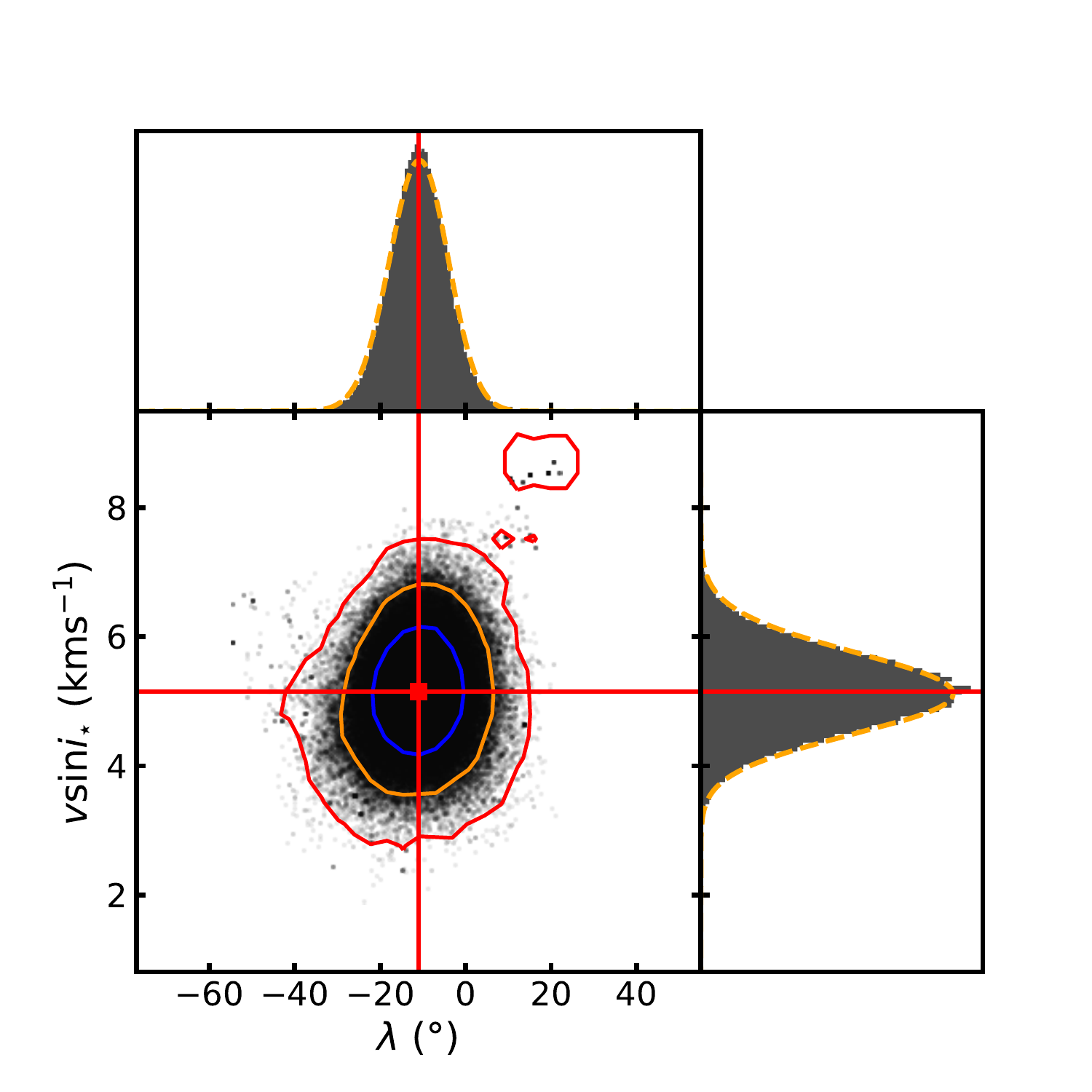}
    \caption{Posterior distributions of the projected spin-orbit angle ($\lambda$) and projected stellar rotational velocity ($v\sin i_{\star}$) for K2-232 from the Markov Chain Monte Carlo Rossiter-McLaughlin simulation. The posterior points inside the blue, yellow, and red contours lie within the $1\sigma$, $2\sigma$, and $3\sigma$ confidence regions, respectively. We have marginalized over $\lambda$ and $v\sin i_{\star}$ and fit each with a Gaussian. The red square denotes the preferred model solution for $\lambda$ and $v\sin i_{\star}$ as given in Table~\ref{table:results}.}
    \label{fig:corner_plot}
\end{figure}

K2-232 was observed with the Automated Planet Finder (APF) telescope \citep{Vogt2014} on UT 29 October 2019 and UT 04 January 2020,
covering two transit events of the warm Jupiter K2-232\,b. Each transit was observed as a series of 20-minute exposures using the APF's $1\times 3\arcsec$ slit, which produces a typical spectral resolution of 90{,}000. The APF uses the iodine-cell technique
for radial-velocity determination: a container
of gaseous iodine is placed in the converging beam of the telescope, imprinting the 5000--6200\,\AA\ region of the incoming stellar spectra with a dense forest of absorption lines that acts as a wavelength calibrator and provides a means of determining the spectrometer's point-spread function (PSF). The October exposures have an average of $2630$ counts/pixel in the iodine region of the spectrum, while the January exposures have a lower average of $1494$ counts/pixel due to a combination of clouds and worse atmospheric seeing ($1.52''$ and $1.98''$ for October and January data, respectively).

%The October exposures have an average of $2630$ counts/pixel in the iodine region of the spectrum \textbf{and an average seeing of $1.52''$}, while the January exposures have a lower average of $1494$ counts/pixel \textbf{and an average seeing of $1.98''$} due to the presence of clouds.

Once the iodine region of the spectrum has been extracted, it is divided into 2\,\AA\ chunks. Each chunk is analyzed using the spectral synthesis technique originally described by \citet{Butler1996}, which disentangles the stellar spectrum from the iodine absorption lines and produces an independent measure of the wavelength, instrument PSF, and Doppler shift. The Doppler velocity is calculated as the weighted mean of the velocities that are determined from each of the $\approx$700 chunks. The final internal uncertainty of each velocity is the standard deviation of the mean of all 700 chunk velocities. The October and January exposure series have mean internal uncertainties of 4.44 and 8.20 m\,s$^{-1}$, respectively. The radial velocities with uncertainties can be found in Table~\ref{RM data}.

\section{Spin-Orbit Angle Determination}

\begin{figure*}
   \includegraphics[width = 1.0\textwidth]{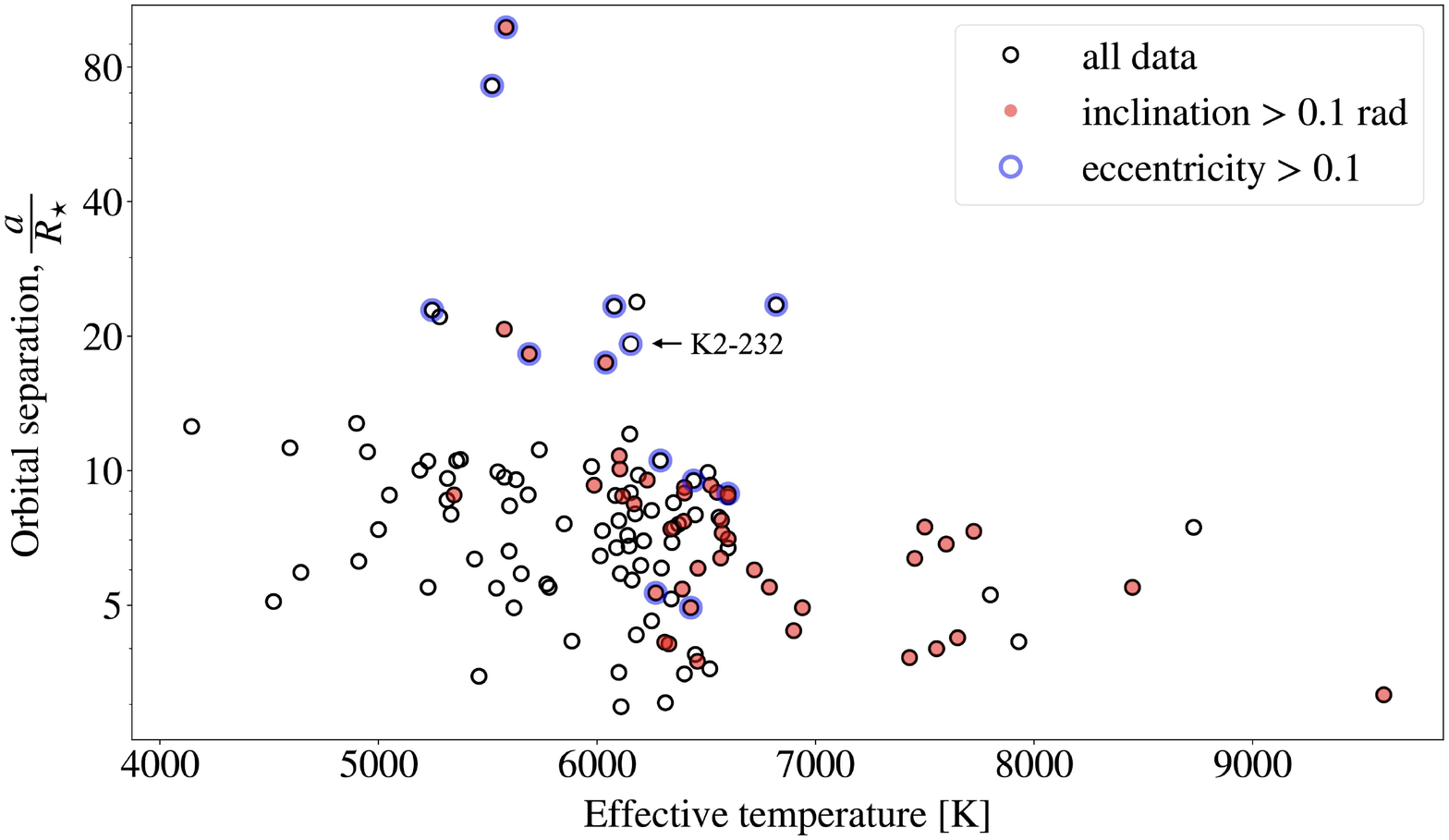}
    \caption{Key variables that may affect tidal dissipation rates, provided
    for all the transiting giant planets
    ($M_{\rm p} > 0.3\,M_{\rm Jup}$)
    for which the Rossiter-McLaughlin effect has been reported.
    The horizontal axis
    shows effective temperature of the host star, and
    the vertical axis shows the dimensionless orbital separation
    (semimajor axis divided by stellar radius). 
    Red points indicate the sky-projected obliquity
    exceeds 0.1~rad (5.7 deg), and blue encircled points
    indicate the orbital
    eccentricity exceeds 0.1, in both
    cases with at least 3-$\sigma$ confidence.
    The scarcity of misaligned or eccentric
    systems with $a/R_\star \lesssim 10$ and $T_{\rm eff} \lesssim 6000\,{\rm K}$ suggests that tidal effects have damped
    the obliquity and eccentricity of at least some of those systems.
    }
    \label{fig:tides}
\end{figure*}

We determined the sky-projected spin-orbit angle ($\lambda$) for K2-232\,b using the \texttt{Allesfitter} code\footnote{https://github.com/MNGuenther/allesfitter} \citep{Max2020}. We simultaneously modeled the \textit{K2} photometry of the target; the in-transit APF radial-velocity time series (which exhibit
the Rossiter-McLaughlin effect); and the out-of-transit radial velocities available from the discovery papers \citep{Yu2018, Brahm2018}, taken with the APF, FEROS, CORALIE, and HARPS spectrographs.

The model parameters included the orbital period ($P$), time
of mid-transit at a reference epoch ($T_{0}$), cosine of the orbital inclination ($\cos{i}$), planet-to-star radius ratio ($R_{P}/R_{\star}$), sum of radii divided by the orbital semi-major axis ($(R_{\star}+R_{P})/a$), radial-velocity semi-amplitude ($K$), parameterized eccentricity and argument of periastron ($\sqrt{e}\,\cos{\,\omega}$, $\sqrt{e}\,\sin{\,\omega}$), quadratic limb darkening coefficients ($q_{1}$, $q_{2}$), sky-projected spin-orbit angle ($\lambda$), and sky-projected stellar rotational velocity ($v\sin i_{\star}$). 
Uniform priors were adopted for all of these parameters.
Initial guesses for $P$, $T_{0}$, $\cos{i}$, $R_{P}/R_{\star}$, $(R_{\star}+R_{P})/a$, $K$, $\sqrt{e}\,\cos{\,\omega}$, and $\sqrt{e}\,\sin{\,\omega}$ were set to the values reported
by \citet{Yu2018}. The initial guess for each limb darkening coefficient was $0.5$. We included standard instrument offsets between radial velocities obtained from each spectrograph in the fitting. We also placed uniform priors on the additive radial velocity offsets between the
time series obtained on each transit night and the
radial velocities gathered outside of transit on other
nights. 
The additive offsets account for any additional astrophysical or instrumental noise on timescales longer than $\sim6\,$hours but shorter than the total time span of all observations ($2\,$years). 
These priors were bounded by reasonable intervals of $\pm1000$\,\mos. 
%$\lambda$ was allowed to vary between $-180$ and 
%$+180$\,degrees.
Table~\ref{table:results} summarizes the model
parameters, priors, and results.

We conducted three joint fits. First, we fit two separate models to independently obtain a Rossiter-McLaughlin measurement from each of the two transit events. Then, we modeled the combined dataset. Together with the in-transit data, each fit incorporated the $K2$ photometry and out-of-transit radial velocity data. For each fit, we sampled the posterior distributions of the model parameters using the Metropolis-Hastings Markov Chain Monte Carlo (MCMC) algorithm with 100 independent walkers each with 300{,}000 total accepted steps.  The results, as listed in Table~\ref{table:results}, provided solutions for transit and radial velocity parameters, as well as the sky-projected spin-orbit angle $\lambda$, the sky-projected stellar rotational velocity $v\sin i_{\star}$, and the associated $1\sigma$ uncertainties. The radial-velocity and transit parameters obtained from our analysis are in good agreement with the values from \citet{Brahm2018} and \citet{Yu2018}.

The observations and the resulting best-fit model for the phased transit are shown in Figure~\ref{fig:combined_transit}, and Figure~\ref{fig:corner_plot} shows the posterior distributions of $\lambda$ and $v\sin i_{\star}$. Figure~\ref{fig:corner_plot} reveals a nearly ideal 2D Gaussian distribution, suggesting that $\lambda$ and $v\sin i_{\star}$ are not strongly correlated with each other. This is generally the case for transiting systems with moderate impact parameters ($a\cos i/R_{\star}\approx 0.5$) as is the case for K2-232\,b \citep{GaudiWinn2007}. The small fractional uncertainty in the $v\sin i_{\star}$ prior ($\sim12\%$) further helps in determining $\lambda$.

We find the best-fit projected spin-orbit angle and projected stellar rotational velocity from the phased transit to be $\lambda = -11.1\pm6.6^{\circ}$ and $v\sin i_* = 5.15\pm0.62$\,\kms, which agree with the corresponding values from independent fits to the two transit events. Our results suggest that the orbit of K2-232\,b is aligned with the spin axis of its host star. 

\section{Discussion}

Previous studies have indicated that obliquity excitation is not
specific to hot Jupiters: giant planets with moderately wider orbits --- wide
enough to expect tidal dissipation to be insignificant --- have also been
observed with large orbital inclinations relative to the stellar equator. However, the number of warm Jupiters with measured spin-orbit angles is small, meaning that it is not yet clear how prevalent these large misalignments are across the longer-period giant planet population. The low obliquity of K2-232\,b provides an example of a relatively well-aligned system among the longer-period giant planets. Additional observations are necessary to draw robust conclusions regarding the distribution of spin-orbit angles for warm Jupiters.

While K2-232\,b is well-aligned with its host star, the planet's orbit is moderately eccentric.
One might expect eccentricity and inclination to be coupled,
in the sense that many interactions that would
excite eccentricity would also excite inclination, and
vice versa. Some examples of such interactions are planet-planet perturbations
and Lidov-Kozai oscillations. Among the observations, we sometimes see orbits with
high eccentricities and high inclinations, such as HD 80606\,b \citep{Winn2009} 
and WASP-8\,b \citep{Queloz2010}.  
There are also cases of low eccentricities and high inclinations, such
as HAT-P-7\,b \citep{Winn2009b}, CoRoT-1\,b \citep{Pont2010}, and KELT-9 \citep{Gaudi2017}.
And, there are cases of high eccentricities
and low inclinations (at least as projected on the sky) similar to K2-232\,b, such as
HAT-P-2\,b \citep{Winn2007}, HD 17156\,b \citep{Narita2009}, and HAT-P-34\,b \citep{Albrecht2012}.

The absence of a firm relationship between inclination and eccentricity
is a clue that there are multiple physical processes at play, some of which are
specific to eccentricity or to inclination.
For examples, disk–planet interactions almost always damp inclinations, but they can result in a growth of eccentricity under the action of Lindblad resonances \citep{Goldreich2003}. As for the reverse, some secular effects, such as nodal precession due to an external inclined perturber, can modify inclination without necessarily modifying eccentricity \citep{Innanen1997}. More measurements may be needed to delineate the relative contributions of each effect.

One pattern that is already clear is that the
closest-orbiting planets ($a/R_{\star} \lesssim 10$) around
cool stars ($T_{\rm{eff}} \lesssim$ 6{,}000\,K)
tend to have both circular and well-aligned orbits.
Figure~\ref{fig:tides} shows $a/R_\star$ versus stellar $T_{\rm eff}$
for all of the transiting giant planets for which $\lambda$ has
been reported based on observations of the Rossiter-McLaughlin
effect. The data were drawn from the TEPCAT
database\footnote{https://www.astro.keele.ac.uk/jkt/tepcat/} and restricted
to cases for which the planet is more massive
than Saturn (0.3\,$M_{\rm Jup}$).
The blue encircled points are those systems for which Doppler observations
have shown that the eccentricity exceeds 0.1 and
differs from zero with at least 3-$\sigma$ confidence.
The red filled points are those
for which Rossiter-McLaughlin observations
have shown that $\lambda$ exceeds 0.1 radians (5.7$^{\circ}$)
and differs from zero with at least 3-$\sigma$ confidence.

\begin{deluxetable}{lccccc}
\tablecaption{Radial velocities for the K2-232 system collected with the APF in this work \label{RM data}}
\tablehead{\colhead{Time (BJD)} & \colhead{RV (\mos)} & \colhead{$\sigma_{\rm RV}$ (\mos)}}
\startdata
% \smallskip                      \\
\multicolumn{3}{c}{Oct. 29, 2019}\\
% \smallskip\\
2458785.76135 & 54.43 & 6.22 \\
2458785.77494 & 58.95 & 4.97 \\
2458785.78924 & 46.95 & 4.93 \\
2458785.80668 & 51.24 & 5.0 \\
2458785.82189 & 40.96 & 4.69 \\
2458785.83645 & 36.59 & 4.67 \\
2458785.85075 & 17.47 & 4.15 \\
2458785.86519 & 19.14 & 4.21 \\
2458785.8791 & 0.0 & 4.39 \\
2458785.89446 & 17.18 & 4.55 \\
2458785.90838 & 9.12 & 4.32 \\
2458785.92253 & 10.52 & 4.06 \\
2458785.93707 & 22.14 & 4.04 \\
2458785.95148 & 25.33 & 3.95 \\
2458785.96582 & 19.29 & 4.56 \\
2458785.98007 & 24.96 & 3.62 \\
2458785.99483 & 20.63 & 4.26 \\
2458786.00914 & 31.08 & 4.38 \\
2458786.02352 & 23.42 & 4.19 \\
2458786.03779 & 3.06 & 4.16 \\
2458786.05224 & 30.56 & 4.07 \\
2458786.06682 & 13.6 & 4.18 \\
\hline
% \smallskip                      \\
\multicolumn{3}{c}{Jan. 4, 2020}\\
% \smallskip\\
2458852.62955 & 9.26 & 7.32 \\
2458852.64381 & 5.75 & 10.65 \\
2458852.65918 & 12.45 & 6.25 \\
2458852.67522 & 10.99 & 5.22 \\
2458852.68760 & -1.73 & 5.16 \\
2458852.70191 & 4.28 & 4.99 \\
2458852.71790 & 3.62 & 4.65 \\
2458852.73021 & -3.69 & 4.68 \\
2458852.74530 & 5.81 & 4.29 \\
2458852.76125 & 20.99 & 4.89 \\
2458852.77514 & 18.67 & 5.77 \\
2458852.78896 & 10.38 & 5.71 \\
2458852.80286 & 42.77 & 7.80 \\
2458852.81951 & 21.92 & 5.60 \\
2458852.83234 & 29.54 & 4.96 \\
2458852.84758 & 12.73 & 6.02 \\
2458852.86249 & -19.92 & 8.31 \\
2458852.87701 & -34.37 & 9.8 \\
\enddata
\end{deluxetable}

\begin{deluxetable*}{lccccc}
%\tablenum{1}
\tablewidth{800pt}
\tabletypesize{\scriptsize}
\tablecaption{System Parameters, Priors, and Results for K2-232\label{table:results}}
\tablehead{
\colhead{Parameter} & \colhead{Priors$\tablenotemark{a}$} & \colhead{Results 29 Oct. 2019} & \colhead{Results 4 Jan. 2020} & \colhead{Preferred Solution} \\
\colhead{} & \colhead{} & \colhead{} & \colhead{} & \colhead{Results combined transits}}
\startdata
Fitted Parameters:\\
Orbital period, $P$ (days) &                                                $ 10; 11.168454; 12 $             & $11.168452\pm0.000025$                   & $11.168466\pm0.000024$                  & $11.168455\pm0.000023$             \\
Mid-transit epoch (2450000-BJD),    $T_{0}$ &                               $ 7825.3\tablenotemark{b} $       & $8305.59490\pm0.00099$                   & $8339.1008\pm0.0010$                    & $8339.10040\pm0.0010$              \\
Cosine of the orbital inclination, $\cos{i}$  &                             $ 0; 0.015; 1 $                   & $0.0345_{-0.0024}^{+0.0028}$             & $0.0344_{-0.0024}^{+0.0029}$            & $0.0344_{-0.0023}^{+0.0028}$       \\
Planet-to-star radius ratio, $R_{P}/R_{\star}$ &                            $ 0; 0.08868; 1 $\                & $0.09142_{-0.00049}^{+0.00052}$          & $0.09138_{-0.00048}^{+0.00052}$         & $0.09140_{-0.00048}^{+0.00051}$    \\
Sum of radii divided by the orbital semimajor axis, $(R_{\star}+R_{P})/a$ & $ 0; 0.05654; 1 $\                & $0.0676_{-0.0023}^{+0.0026}$             & $0.0676_{-0.0023}^{+0.0027}$            & $0.0675_{-0.0022}^{+0.0026}$       \\
RV semi-amplitude, $K$ ($\rm m \ s^{-1}$) &                                 $ 0; 33; 1000 $\                  & $32.4\pm2.5$                             & $32.3\pm 2.6$                           & $32.4\pm2.5$                       \\
Eccentricity parameter 1, $\sqrt{e}$ cos$\omega$  &                         $ -1.0; -0.453; 1.0 $\            & $-0.441_{-0.030}^{+0.035}$               & $-0.440_{-0.031}^{+0.036}$              & $-0.440_{-0.030}^{+0.035}$         \\
Eccentricity parameter 2, $\sqrt{e}$ sin$\omega$ &                          $ -1.0; -0.231; 1.0 $\            & $-0.217_{-0.061}^{+0.076}$               & $-0.217_{-0.062}^{+0.081}$              & $-0.221_{-0.060}^{+0.074}$         \\
Limb-darkening coefficient 1, $q_{1}$ &                                     $ 0; 0.5; 1.0 $                   & $0.538\pm0.068$                          & $0.544\pm0.067$                         & $0.540\pm0.068$                    \\
Limb-darkening coefficient 2, $q_{2}$ &                                     $ 0; 0.5; 1.0 $\                  & $0.187_{-0.043}^{+0.054}$                & $0.183_{-0.041}^{+0.054}$               & $0.185_{-0.043}^{+0.054}$          \\
Stellar rotation velocity, $v\sin i_{\star}$ ($\rm km \ s^{-1}$) &          $ 0; 5; 10 $                      & $5.23\pm0.72$                            & $5.5_{-1.6}^{+1.8}$                     & $5.15\pm0.62$                      \\
Projected spin-orbit angle, $\lambda$ (deg)  &                              $ -180; 0; 180 $                  & $-14.5\pm7.7$                            & $-2_{-16}^{+12}$                        & $-11.1\pm6.6$                      \\
Relative RV Offset for in-transit APF, Oct. 29 2019 (\mos)                 & $ -1000; 0; 1000$                              & $32.0\pm3.1$                        & ...                                     & $32.8\pm2.8$ \\
Relative RV Offset for in-transit APF, Jan. 4 2020  (\mos)                  & $ -1000; 0; 1000$                              & ...                                      & $8.6\pm4.3$                       & $7.1_{-3.5}^{+3.2}$\\
Relative RV Offset for out-of-transit APF  (\mos)                   & $ -1000; 0; 1000$                              & $-98.7\pm1.9$                  & $98.7_{-1.8}^{+1.9}$          & $-98.7_{-1.7}^{+1.9}$\\
Relative RV Offset for Coralie  (\mos)                 & $ -1000; 0; 1000$                            & $-57_{-22}^{+19}$   & $-57_{-22}^{+19}$             & $-58_{-22}^{+19}$\\
Relative RV Offset for FEROS  (\mos)                   & $ -1000; 0; 1000$                              & $3.5\pm2.1$            & $3.4\pm2.1$                      & $3.5\pm2.1$\\
Relative RV Offset for HARPS  (\mos)                  & $ -1000; 0; 1000$                              & $-0.7\pm2.7$           & $-0.6_{-2.5}^{+2.7}$          & $-0.8_{-2.5}^{+2.7}$\\
\\
Derived Parameters:\\
Planetary radius  $R_\mathrm{b}$ ($\mathrm{R_{jup}}$)            &...                                         & $0.380_{-0.044}^{+0.049}$     & $0.378_{-0.045}^{+0.049}$   & $0.378_{-0.045}^{+0.047}$          \\
Planetary Mass    $M_\mathrm{b}$ ($\mathrm{M_{jup}}$)            &...                                         & $1.097\pm0.025$               & $1.096\pm0.025$             & $1.097\pm0.024$             \\
Impact parameter   $b$                                           &...                                         & $0.587\pm0.014$               & $0.586\pm0.013$             & $0.587\pm0.013$          \\
Transit duration  $T_\mathrm{14}$   (h)                          &...                                         & $5.2893\pm0.010$              & $5.2893\pm0.010$            & $5.2895\pm0.0099$          \\
Transit depth     $\delta$                                        &...                                        & $0.009331\pm0.000019$         & $0.009330\pm0.000018$       & $0.009330_{-0.000021}^{+0.000019}$    \\
Inclination $i$ ($^{\circ}$)               &...                                                               & $88.02_{-0.16}^{+0.14}$                  & $88.03_{-0.17}^{+0.14}$                 & $88.03_{-0.16}^{+0.13}$            \\
Eccentricity $e$                           &...                                                               & $0.245_{-0.022}^{+0.020}$                & $0.246_{-0.022}^{+0.020}$               & $0.247_{-0.021}^{+0.020}$          \\
Argument of periastron $\omega$ (deg)      &...                                                               & $206.3_{-9.4}^{+8.0}$                    & $206.3_{-9.8}^{+7.8}$                   & $206.7_{-9.0}^{+7.8}$              \\
Limb darkening; $u_\mathrm{1}$             &...                                                               & $0.276_{-0.053}^{+0.061}$                & $0.272_{-0.052}^{+0.058}$               & $0.273_{-0.053}^{+0.059}$          \\
Limb darkening; $u_\mathrm{2}$             &...                                                               & $0.461_{-0.10}^{+0.098}$                 & $0.468_{-0.10}^{+0.096}$                & $0.464_{-0.10}^{+0.097}$           \\
\enddata
\tablenotetext{a}{The uniform priors are presented in the form of three numbers: the first number is the lower bound, the middle number is the initial guess, and the last number is the upper bound.}
\tablenotetext{b}{We provided a reference mid-transit epoch for $T_{0}$. During the fit, \texttt{Allesfitter} can shift epochs to the data center to derive an optimal $T_{0}$.}
\end{deluxetable*}

Figure~\ref{fig:tides} shows a mixture of red points, blue points, red and blue
points, and uncolored points, illustrating the point
that eccentricity and inclination are not always observed
to go together.  The subject of this paper, K2-232, is represented
by a blue point with
$a/R_\star\approx 19$ and $T_{\rm eff} \approx 6154$\,K.

Note, too, that the red and blue points
are almost \textit{all} found outside of the lower left corner.
The systems with $a/R_\star\lesssim 10$ and $T_{\rm eff} \lesssim 6000$\,K
are not demonstrably inclined or eccentric (although in some cases
the uncertainties are large).
Apparently, the hottest Jupiters around relatively cool
stars have lower eccentricities and inclinations than the other
members of the sample.
This may be evidence for tidal dissipation, which acts
to dynamically cool the system. Dissipation rates
are strong functions of orbital separation and may also
depend sensitively on the internal structure of the star
(although for eccentricity damping, tidal dissipation within the planet
is expected to be at least as important the dissipation within
the star).

Thus, the current dataset, limited though it might be,
provides evidence that tidal dissipation has reoriented
and circularized the orbits of hot Jupiters.

%\software{MultiNest \citep{Feroz}, Forecaster \citep{Chen2017}}

\acknowledgements

The authors thank Maximilian N. G${\rm \ddot u}$enther for helpful discussions on the \texttt{Allesfitter} and
are also grateful to the anonymous referee for their constructive comments and suggestions. J.N.W.\ thanks the Heising-Simons Foundation for support. M.R. is supported by the National Science Foundation Graduate Research Fellowship Program under Grant Number DGE-1752134.
We are grateful to Metrics for providing helpful suggestions on the figure design.
Part of this research was carried out at the Jet Propulsion Laboratory,
California Institute of Technology, under a contract with the National
Aeronautics and Space Administration (80NM0018D0004). This work is supported by China National Astronomical Data Center (NADC), Chinese Virtual Observatory (China-VO), Astronomical Big Data Joint Research Center, co-founded by National Astronomical Observatories, Chinese Academy of Sciences and Alibaba Cloud.

\bibliography{main}

\begin{thebibliography}{}
\expandafter\ifx\csname natexlab\endcsname\relax\def\natexlab#1{#1}\fi
\providecommand{\url}[1]{\href{#1}{#1}}

\bibitem[{{Albrecht} {et~al.}(2012){Albrecht}, {Winn}, {Johnson}, {Howard},
  {Marcy}, {Butler}, {Arriagada}, {Crane}, {Shectman}, {Thompson}, {Hirano},
  {Bakos}, \& {Hartman}}]{Albrecht2012}
{Albrecht}, S., {Winn}, J.~N., {Johnson}, J.~A., {et~al.} 2012, \apj, 757, 18

\bibitem[{{Bate} {et~al.}(2010){Bate}, {Lodato}, \& {Pringle}}]{Bate2010}
{Bate}, M.~R., {Lodato}, G., \& {Pringle}, J.~E. 2010, \mnras, 401, 1505

\bibitem[{{Batygin} {et~al.}(2011){Batygin}, {Morbidelli}, \&
  {Tsiganis}}]{Batygin2011}
{Batygin}, K., {Morbidelli}, A., \& {Tsiganis}, K. 2011, \aap, 533, A7

\bibitem[{{Brahm} {et~al.}(2018){Brahm}, {Espinoza}, {Jord{\'a}n}, {Rojas},
  {Sarkis}, {D{\'\i}az}, {Rabus}, {Drass}, {Lachaume}, {Soto}, {Jenkins},
  {Jones}, {Henning}, {Pantoja}, \& {Vu{\v{c}}kovi{\'c}}}]{Brahm2018}
{Brahm}, R., {Espinoza}, N., {Jord{\'a}n}, A., {et~al.} 2018, \mnras, 477, 2572

\bibitem[{{Butler} {et~al.}(1996){Butler}, {Marcy}, {Williams}, {McCarthy},
  {Dosanjh}, \& {Vogt}}]{Butler1996}
{Butler}, R.~P., {Marcy}, G.~W., {Williams}, E., {et~al.} 1996, \pasp, 108, 500

\bibitem[{{de Laplace}(1796)}]{Laplace1796}
{de Laplace}, P.~S. 1796, {Exposition du syst{\`e}me du monde},
  doi:10.3931/e-rara-497

\bibitem[{Dong {et~al.}(2013)Dong, Katz, \& Socrates}]{dong2013warm}
Dong, S., Katz, B., \& Socrates, A. 2013, The Astrophysical Journal Letters,
  781, L5

\bibitem[{{Fabrycky} \& {Tremaine}(2007)}]{Fabrycky2007}
{Fabrycky}, D., \& {Tremaine}, S. 2007, \apj, 669, 1298

\bibitem[{{Ford} \& {Rasio}(2008)}]{Ford2008}
{Ford}, E.~B., \& {Rasio}, F.~A. 2008, \apj, 686, 621

\bibitem[{{Gaudi} \& {Winn}(2007)}]{GaudiWinn2007}
{Gaudi}, B.~S., \& {Winn}, J.~N. 2007, \apj, 655, 550

\bibitem[{{Gaudi} {et~al.}(2017){Gaudi}, {Stassun}, {Collins}, {Beatty},
  {Zhou}, {Latham}, {Bieryla}, {Eastman}, {Siverd}, {Crepp}, {Gonzales},
  {Stevens}, {Buchhave}, {Pepper}, {Johnson}, {Colon}, {Jensen}, {Rodriguez},
  {Bozza}, {Novati}, {D'Ago}, {Dumont}, {Ellis}, {Gaillard}, {Jang-Condell},
  {Kasper}, {Fukui}, {Gregorio}, {Ito}, {Kielkopf}, {Manner}, {Matt}, {Narita},
  {Oberst}, {Reed}, {Scarpetta}, {Stephens}, {Yeigh}, {Zambelli}, {Fulton},
  {Howard}, {James}, {Penny}, {Bayliss}, {Curtis}, {Depoy}, {Esquerdo},
  {Gould}, {Joner}, {Kuhn}, {Labadie-Bartz}, {Lund}, {Marshall}, {McLeod},
  {Pogge}, {Relles}, {Stockdale}, {Tan}, {Trueblood}, \&
  {Trueblood}}]{Gaudi2017}
{Gaudi}, B.~S., {Stassun}, K.~G., {Collins}, K.~A., {et~al.} 2017, \nat, 546,
  514

\bibitem[{{Goldreich} \& {Sari}(2003)}]{Goldreich2003}
{Goldreich}, P., \& {Sari}, R. 2003, \apj, 585, 1024

\bibitem[{{G{\"u}nther} \& {Daylan}(2020)}]{Max2020}
{G{\"u}nther}, M.~N., \& {Daylan}, T. 2020, arXiv e-prints, arXiv:2003.14371

\bibitem[{{Innanen} {et~al.}(1997){Innanen}, {Zheng}, {Mikkola}, \&
  {Valtonen}}]{Innanen1997}
{Innanen}, K.~A., {Zheng}, J.~Q., {Mikkola}, S., \& {Valtonen}, M.~J. 1997,
  \aj, 113, 1915

\bibitem[{{Kant}(1755)}]{Kant1755}
{Kant}, I. 1755, {Allgemeine Naturgeschichte und Theorie des Himmels}

\bibitem[{{Lai} {et~al.}(2011){Lai}, {Foucart}, \& {Lin}}]{Lai2011}
{Lai}, D., {Foucart}, F., \& {Lin}, D. N.~C. 2011, \mnras, 412, 2790

\bibitem[{McLaughlin(1924)}]{mclaughlin1924some}
McLaughlin, D. 1924, The Astrophysical Journal, 60

\bibitem[{{Naoz}(2016)}]{Naoz2016}
{Naoz}, S. 2016, \araa, 54, 441

\bibitem[{{Narita} {et~al.}(2009){Narita}, {Hirano}, {Sato}, {Winn}, {Suto},
  {Turner}, {Aoki}, {Tamura}, \& {Yamada}}]{Narita2009}
{Narita}, N., {Hirano}, T., {Sato}, B., {et~al.} 2009, \pasj, 61, 991

\bibitem[{{Petrovich}(2015)}]{Petrovich2015}
{Petrovich}, C. 2015, \apj, 799, 27

\bibitem[{{Pont} {et~al.}(2010){Pont}, {Endl}, {Cochran}, {Barnes}, {Sneden},
  {MacQueen}, {Moutou}, {Aigrain}, {Alonso}, {Baglin}, {Bouchy}, {Deleuil},
  {Fridlund}, {H{\'e}brard}, {Hatzes}, {Mazeh}, \& {Shporer}}]{Pont2010}
{Pont}, F., {Endl}, M., {Cochran}, W.~D., {et~al.} 2010, \mnras, 402, L1

\bibitem[{{Queloz} {et~al.}(2010){Queloz}, {Anderson}, {Collier Cameron},
  {Gillon}, {Hebb}, {Hellier}, {Maxted}, {Pepe}, {Pollacco}, {S{\'e}gransan},
  {Smalley}, {Triaud}, {Udry}, \& {West}}]{Queloz2010}
{Queloz}, D., {Anderson}, D.~R., {Collier Cameron}, A., {et~al.} 2010, \aap,
  517, L1

\bibitem[{{Rasio} \& {Ford}(1996)}]{Rasio1996}
{Rasio}, F.~A., \& {Ford}, E.~B. 1996, Science, 274, 954

\bibitem[{{Rogers} {et~al.}(2012){Rogers}, {Lin}, \& {Lau}}]{Rogers2012}
{Rogers}, T.~M., {Lin}, D.~N.~C., \& {Lau}, H.~H.~B. 2012, \apjl, 758, L6

\bibitem[{Rossiter(1924)}]{rossiter1924detection}
Rossiter, R. 1924, The Astrophysical Journal, 60

\bibitem[{{Souami} \& {Souchay}(2012)}]{Souami2012}
{Souami}, D., \& {Souchay}, J. 2012, \aap, 543, A133

\bibitem[{{Southworth}(2011)}]{Southworth2011}
{Southworth}, J. 2011, \mnras, 417, 2166

\bibitem[{{Storch} {et~al.}(2014){Storch}, {Anderson}, \& {Lai}}]{Storch2014}
{Storch}, N.~I., {Anderson}, K.~R., \& {Lai}, D. 2014, Science, 345, 1317

\bibitem[{{Vogt} {et~al.}(2014){Vogt}, {Radovan}, {Kibrick}, {Butler},
  {Alcott}, {Allen}, {Arriagada}, {Bolte}, {Burt}, {Cabak}, {Chloros},
  {Cowley}, {Deich}, {Dupraw}, {Earthman}, {Epps}, {Faber}, {Fischer}, {Gates},
  {Hilyard}, {Holden}, {Johnston}, {Keiser}, {Kanto}, {Katsuki}, {Laiterman},
  {Lanclos}, {Laughlin}, {Lewis}, {Lockwood}, {Lynam}, {Marcy}, {McLean},
  {Miller}, {Misch}, {Peck}, {Pfister}, {Phillips}, {Rivera}, {Sand ford},
  {Saylor}, {Stover}, {Thompson}, {Walp}, {Ward}, {Wareham}, {Wei}, \&
  {Wright}}]{Vogt2014}
{Vogt}, S.~S., {Radovan}, M., {Kibrick}, R., {et~al.} 2014, \pasp, 126, 359

\bibitem[{{Winn} {et~al.}(2010){Winn}, {Fabrycky}, {Albrecht}, \&
  {Johnson}}]{Winn2010}
{Winn}, J.~N., {Fabrycky}, D., {Albrecht}, S., \& {Johnson}, J.~A. 2010, \apjl,
  718, L145

\bibitem[{{Winn} {et~al.}(2009{\natexlab{a}}){Winn}, {Johnson}, {Albrecht},
  {Howard}, {Marcy}, {Crossfield}, \& {Holman}}]{Winn2009b}
{Winn}, J.~N., {Johnson}, J.~A., {Albrecht}, S., {et~al.} 2009{\natexlab{a}},
  \apjl, 703, L99

\bibitem[{{Winn} {et~al.}(2007){Winn}, {Johnson}, {Peek}, {Marcy}, {Bakos},
  {Enya}, {Narita}, {Suto}, {Turner}, \& {Vogt}}]{Winn2007}
{Winn}, J.~N., {Johnson}, J.~A., {Peek}, K. M.~G., {et~al.} 2007, \apjl, 665,
  L167

\bibitem[{{Winn} {et~al.}(2009{\natexlab{b}}){Winn}, {Howard}, {Johnson},
  {Marcy}, {Gazak}, {Starkey}, {Ford}, {Col{\'o}n}, {Reyes}, {Nortmann},
  {Dreizler}, {Odewahn}, {Welsh}, {Kadakia}, {Vanderbei}, {Adams}, {Lockhart},
  {Crossfield}, {Valenti}, {Dantowitz}, \& {Carter}}]{Winn2009}
{Winn}, J.~N., {Howard}, A.~W., {Johnson}, J.~A., {et~al.} 2009{\natexlab{b}},
  \apj, 703, 2091

\bibitem[{{Wu} \& {Lithwick}(2011)}]{Wu2011}
{Wu}, Y., \& {Lithwick}, Y. 2011, \apj, 735, 109

\bibitem[{{Wu} \& {Murray}(2003)}]{Wu2003}
{Wu}, Y., \& {Murray}, N. 2003, \apj, 589, 605

\bibitem[{{Yu} {et~al.}(2018){Yu}, {Rodriguez}, {Eastman}, {Crossfield},
  {Shporer}, {Gaudi}, {Burt}, {Fulton}, {Sinukoff}, {Howard}, {Isaacson},
  {Kosiarek}, {Ciardi}, {Schlieder}, {Penev}, {Vanderburg}, {Stassun},
  {Bieryla}, {Butler}, {Berlind}, {Calkins}, {Esquerdo}, {Latham}, {Murawski},
  {Stevens}, {Petigura}, {Kreidberg}, \& {Bristow}}]{Yu2018}
{Yu}, L., {Rodriguez}, J.~E., {Eastman}, J.~D., {et~al.} 2018, \aj, 156, 127

\end{thebibliography}



\end{document}